\documentclass[11pt,aps,preprint,epsfig,showpacs]{revtex4}
\usepackage{graphicx}
\usepackage{epsfig}
\usepackage{color}

\begin{document}
\title{Stretching of a single-stranded DNA: Evidence for structural transition}
\author{Garima Mishra, Debaprasad Giri${^1}$ and Sanjay Kumar}
\affiliation{Department of Physics, Banaras Hindu University,
Varanasi 221 005, India \\
${^1}$Department of Applied physics, Institute of Technology, 
Banaras Hindu University, Varanasi 221 005, India} 

\begin{abstract}
Recent experiments have shown that the force-extension (F-x) 
curve for single-stranded DNA (ssDNA) consisting only of adenine [poly(dA)] is 
significantly different from thymine [poly(dT)].
Here, we show that the base stacking interaction is not 
sufficient to describe the F-x curves as seen in the experiments. 
A reduction in the reaction co-ordinate arising from the 
formation of helix at low forces and an increase in the distance 
between consecutive phosphates of unstacked bases in the 
stretched state at high force in the proposed
model, qualitatively reproduces the experimentally observed 
features.  The multi-step plateau  in the F-x curve is a 
signature of structural change in ssDNA. 
\end{abstract}
\maketitle

The last ten  years have witnessed a revolution in 
Single Molecules Force Spectroscopy (SMFS) experiments 
involving the manipulation of single biomolecules. 
These experiments allowed detecting inter and intra   
molecular forces and its influence on the properties 
observable at the chain length of biomolecules 
\cite{busta1,rief,cluzel,wang}. 
New challenges have emerged when the semi-microscopic 
changes in the monomer (nucleotide) found to influence 
the elastic property of the single-stranded DNA (ssDNA). Attempts have been 
made to monitor the force-extension (F-x) 
curve \cite{seol2,ke} of RNA and ssDNA consisting of 
only one type of nucleotide. It was found that the 
elastic properties of ssDNA made up with adenine [poly(dA)]
are significantly different from  thymine [poly(dT)] 
(or uracil [poly(rU)]). It is known that base stacking is 
strongest among adenine  and weakest among 
thymine and uracil  \cite{goddard}. 
As a result, base stacking favors  a parallel 
orientation of consecutive bases in poly(dA), but not in 
poly(dT) or poly(rU) \cite{cantor, turner}. Notably, 
theoretical models developed for such studies do not 
include orientation of bases in their 
description \cite{vander,zhou,sain,marko} 
and hence provide a limited picture of the system.

\begin{figure}[t]
\includegraphics [width=3in]{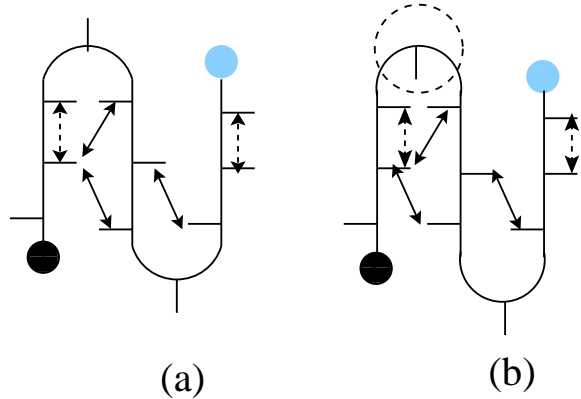}
\caption{(Color online) The schematic representation of 
ssDNA.  Two types of stacking interaction may arise namely 
inter strand shown by the solid line and intra-strand 
shown by dashed line.  In Fig.b we also show the steric 
repulsion among the adjacent bases. The small black circle 
indicates that one end of the ssDNA is kept fixed while 
a force may be applied at the other end (light blue circle).}
\label{fig-1}
\end{figure}

The F-x curves for poly(dT) and poly(rU) show  the effect 
of entropic elasticity, whereas poly(rA) exhibits plateau 
in the F-x curve \cite{seol2} which were found to be absent 
in earlier experiments \cite{busta1,desi,seol1}. 
In a recent experiment, Ke {\it et al.} \cite{ke}
found the existence of multi step plateau in case of poly(dA). 
The first plateau occurs at a force $23 \pm 1$ pN and 
overstretched the nucleotide by $ \sim 74\% $ which has been 
predicted by the model proposed by Buhot and Halperin \cite{buhot}. 
This prediction has also been observed in case of homopolymeric 
RNA \cite{seol2}. 
A plateau obtained in case of ssRNA (poly(rA)) has been 
explained on the basis of seven parameters \cite{seol2} 
in the Zimm-Bragg model \cite{zimm}. The qualitative 
understanding of the first plateau has been achieved in terms 
of the unwinding of helical structure of poly(A) arising due 
to base stacking.  This transition appears to be weakly 
cooperative, but needs further attention from the statistical 
mechanics point of view \cite{buhot,buhot2,seol2}.
Moreover, the second plateau which occurs at a force 
$113 \pm 1$ pN and over stretches ssDNA by an additional 
$16\%$ \cite{ke} was not predicted by the model proposed 
by Buhot and Halperin \cite{buhot}. Though Ke {\it et al} 
conjectured that the second plateau is associated with structural
transition \cite{n3}, but high level {\it ab initio} quantum
mechanical calculation did not support it \cite{ke,folo}.

\begin{figure}[t]
${}$ \hspace {-.2in} \includegraphics[width=2.0in]{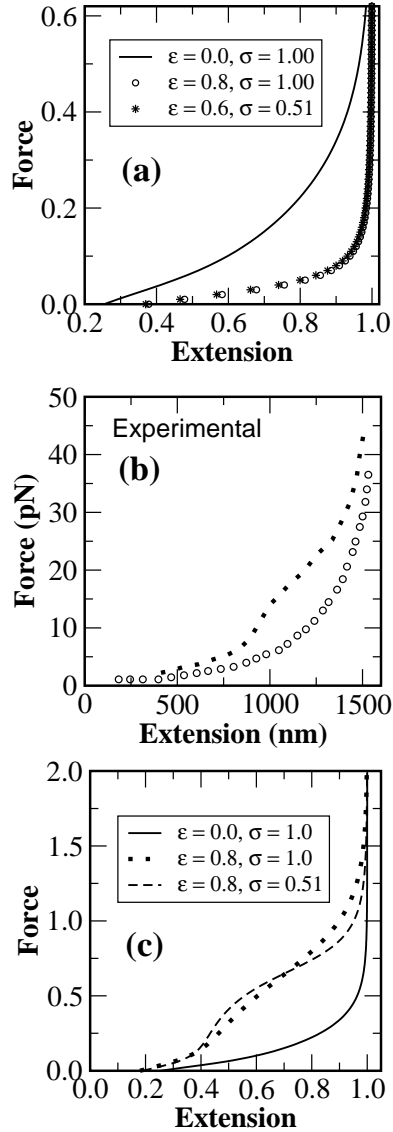}
\caption{(a) F-x curve for the model ssDNA for different values of 
$\epsilon$. $\epsilon = 0 $ corresponds to poly(T) (or poly(U)), 
which shows entropic response \cite{seol2,ke}. $\epsilon \neq 0$ 
represents the case where stacking play a role 
in the formation of helix ({\it e.g.} adenine); 
(b) Experimental F-x curve \cite{seol2} for the poly(rA). 
Here F is in pN. 
(c) Figure shows the F-x curve obtained from the model which 
includes formation of helix (dotted line). In Fig a and c force 
and extension are in the reduced unit.}
\label{fig-3}
\end{figure}

We adopt a more realistic model of DNA where directionality of 
bases has been included apart from excluded volume and non-native
base pairing interaction to study the F-x curve. In order to study 
the consequence of orientation of bases, we allow bases to orient 
along phosphate bonds in the model \cite{turner}. If two bases 
are parallel to each other, we say bases are stacked and associate 
an attractive interaction $\epsilon$ between them. 
The issue of cooperativity can also be studied exactly in this model
by associating a cooperativity factor ($\sigma$) between stacked and 
unstacked domain \cite{zimm}. In this paper, we will show that 
only base stacking and cooperativity factor are not sufficient to 
describe the F-x  curve, but further semi-microscopic modifications 
are needed. Our results based on exact enumeration provide 
unequivocal support that  the multi-step plateau is the result of 
structural changes in ssDNA, a field which warrants further studies.

The model discussed above \cite{kgs} is general and can be 
defined in any dimension, though for computational limitations, 
we restrict ourselves in two dimension. We consider a linear polymer chain 
consisting of either (A) or (T) which are described by 
self-avoiding walks (SAWs) on a square lattice. The bases are 
associated with the link between two monomers as depicted in Fig. 1. 

The present model is a coarse grained one where the microscopic 
distinction between nucleotide A and C  has been ignored. 
Moreover, Seol et al. \cite{seol2} have shown that the 
normalized F-x curve for poly(rA) and poly(rC) have similar 
elastic response. In order to study the multi-step 
plateau \cite{ke},  we consider following cases: 
(i) In absence of the stacking interaction, bases are free 
to orient and the ssDNA (or RNA) behaves like a flexible chain.  
This will correspond to the case of poly(dT) [or poly(rU)], 
(ii) due to the stacking interaction, bases can stack at 
low temperature which may represent the case of  poly(dA) 
[or poly(rA)] \cite{seol2}. 
With these constraints, we enumerate all possible conformations 
of the ssDNA. The partitions function ($Z_N$) of the system 
can be written as 

\begin{eqnarray}
Z_N & = &\sum_{all \; walks} C(N_1,N_2,x) e^{\beta N_1 \epsilon} 
e^{\beta F x} e^{-\beta N_2 \epsilon_w}, 
\end{eqnarray}
where $\beta = \frac{1}{k_B T}$.  Here, $\epsilon_w$, $T$, $F$ 
and $k_B$ are the wall energy, temperature, applied force and 
the Boltzmann constant respectively. $C(N_1,N_2,x)$ is the number 
of distinct conformations of walk of length $N$. $N_1$ and $N_2$ 
are the number of intra stacked bases and the number of walls 
between stacked and unstacked region of the chain whose ends 
are at a distance $x$ apart. $\sigma (= e^{-\beta \epsilon_w}$) 
is the Boltzmann weight for the wall energy which is termed as a 
cooperative factor.
The value of $\sigma$ in between 0.5 to 1 corresponds to weakly 
cooperative, while $\sigma \rightarrow 0$ referred as strongly 
cooperative \cite{zimm}.
Time required to enumerate these conformations increases as $\mu^N 2^N$ 
because of extra degree of freedom associated with bases (2 possible 
orientation) \cite{vander,kj}. 
Here $\mu$ is the connectivity constant of the lattice. The model 
also takes care of steric effect among adjacent bases and conformations 
like the one shown in Fig. 1(b) have not been incorporated in the 
partition function. The quantity of experimental interest {\it i.e } 
the reaction co-ordinate \cite{busta1,buhot} in this case is the 
extension in the chain, which  can be obtained from the expression 
$\langle x \rangle = \frac{1}{Z_N} \sum_{N_1,N_2,x} x C(N_1, N_2, x)  
e^{\beta N_1 \epsilon} e^{\beta F x} 
e^{-\beta N_2 \epsilon_w}$.
 
\begin{figure}[t]
\includegraphics[width=5.in]{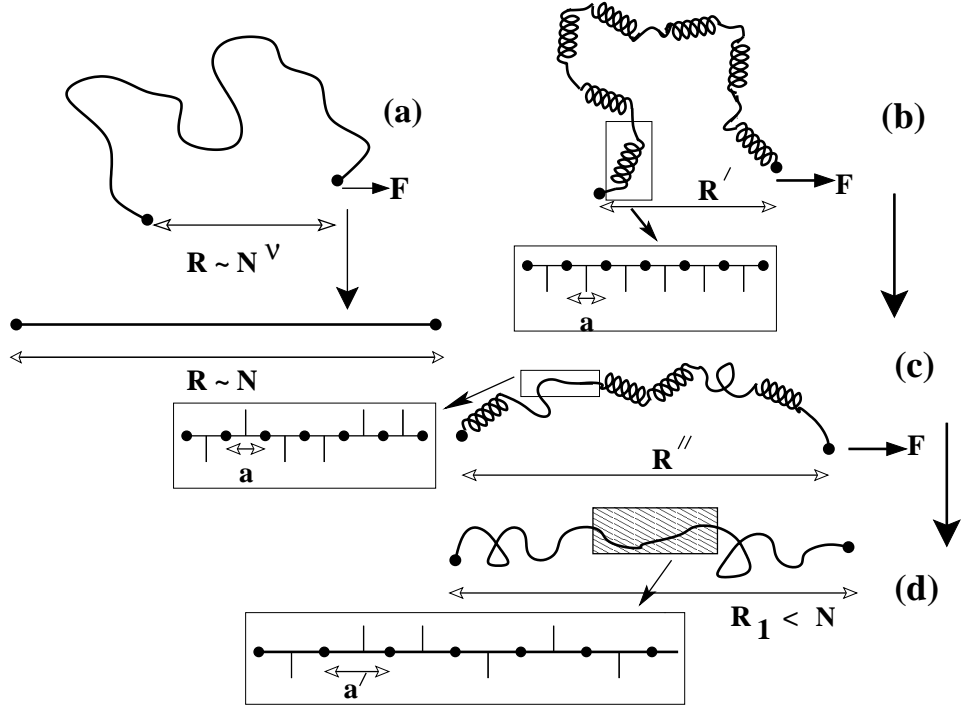} \\
\caption{(a) The Schematic representation of poly(T). 
(b) In case of poly(A), bases can stack and form a helix like 
structure. Inset shows the base stacking for specific segment; 
(c) shows the unwinding of helix where consecutive phosphate 
distance remain same; 
(d) Here the chain is in the stretched state where few bases 
are stacked. Flipping of bases in presence of high force shows 
the extension in the consecutive phosphate {\it i.e} 
$a^{\prime} > a$.}
\label{fig-2}
\end{figure}

\begin{figure}[t]
\hspace {0.0in}\includegraphics[width=2.0in]{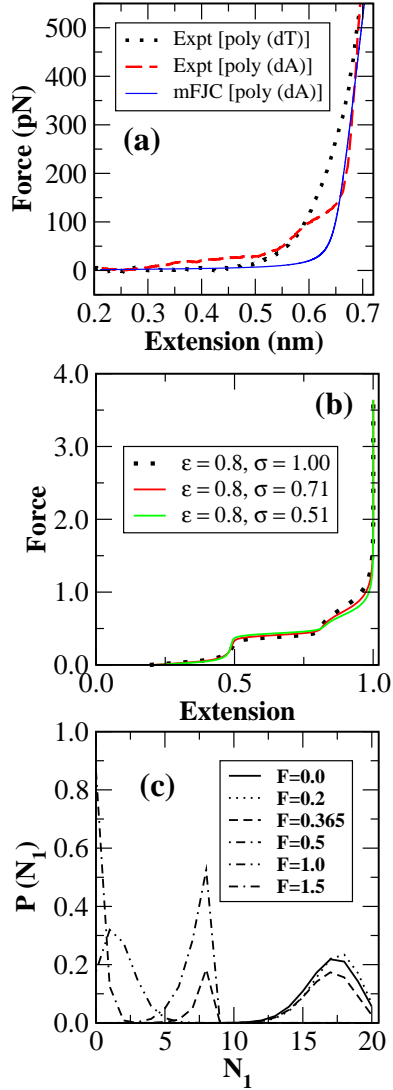} 
\caption{(Color online)(a) Experimental F-x curve \cite{ke} for the poly(dT)
[dotted] and poly(dA) [dashed] of ssDNA. Solid line is for poly(dA)
using mFJC \cite{busta1}; 
(b) Figure shows the F-x curve where the formation of helix 
as well as increase in consecutive phosphate distance between 
bases has been taken into consideration (force and extension 
are in the reduced unit).  One can see that inclusion of such
considerations qualitatively reproduce the F-x curve obtained
by Ke {\it et al} \cite{ke} and exhibits the multi-step plateau;
(c) Figure shows the probability distribution curves for poly(dA).
}
\label{fig-4}
\end{figure}
In the following, we set $k_B = 1$ and plot the F-x curves 
for different values of $\epsilon$ in Fig. 2(a). In absence 
of stacking interaction ($\epsilon =0$), we reproduce the 
entropic response of the chain as seen in case of poly(rU) 
(or poly(dT)) \cite{seol2,ke}. 
In Fig. 2(a) one can see that the curve for poly(rA) shifts 
to the right of poly(rU). Moreover to keep extension say 
at $0.8$, one requires less force for poly(rA) compared 
to poly(rU). In Fig. 2(b), we plot the experimental curve 
obtained by Seol {\it et al.} \cite{seol2}. It is evident 
that curve for poly(rA) shifts to the left of poly(rU) and 
to keep extension say $1200$ nm, one requires almost the 
double force than the poly(rU).  This indicates that apart 
from the stacking interaction and cooperativity there 
must be some other mechanism involved in the elasticity 
of poly(rA). It was proposed \cite{buhot} that since base 
stacking favors formation of helices, the effective 
end-to-end distance ($R'$) decreases as shown in Fig. 3(b). 
Since, we have the exact information about the number of stacked bases 
($N_1$) participating in 
the formation of helix, the reduction in length can be 
calculated exactly in the proposed model.  Following the 
procedure adopted in Ref. \cite{buhot}, we reduce the 
reaction coordinate $x$ by a certain factor proportional 
to $N_1$ {\it i.e.} $x'=x-N_1\alpha$. The value 
of proportionality constant $\alpha =b/a \approx 0.63$ has be obtained by 
setting $a= 5.9 A^o$ and $b=3.7 A^o$ \cite{seol2,buhot}. 
Here $a$ and $b$ are inter phosphate distance and rise 
in length of helix per nucleotide respectively. 
The modified F-x curve, 
with this constraint \cite{ssdna}, has been shown in Fig. 2(c). 
Surprisingly this consideration not only shifts the curve to 
the left as seen in experiment (earlier shifting to the 
right in absence of helix), but also qualitatively 
reproduces the F-x curve obtained by Seol {\it et al} \cite{seol2}. 

In Fig. 2(a), we show that the force-extension 
curve obtained by using cooperativity parameter $\sigma = 0.51$ 
and $\epsilon = 0.6$, which overlaps with $\sigma=1.0$
(no cooperativity) and $\epsilon = 0.8$. Therefore, stacking 
interaction reproduces the F-x curve for the low 
cooperativity. In fact, we find that the nature 
of curve remains almost same upto $\sigma = 0.71$. 
This is in accordance with earlier studies where it was also
found that stacking decreases the cooperativity of melting of
homopolymeric DNA \cite{moro}

When cooperativity increases, flatness of the curve [Fig. 2(c)] 
also increases.  
The reduction in reaction coordinate ($x'=x-N_1\alpha$) along 
with $\sigma$ modifies Eq. (1) which gives rise such effect.  
From Fig.2(c), we also note that the curve with cooperativity ($\sigma = 0.51$)
crosses the curve without cooperativity ($\sigma = 1$) at the extension 
($\approx 0.7$). We identified this as a crossover point where chain 
goes from the extended to the stretched state \cite{kj}.

Existence of the second plateau [Fig. 4(a)] at higher forces as 
seen in case of poly(dA) \cite{ke} requires further refinement
in  the model. At such a high force ($ \sim 113$ pN) chain 
must be in the stretched state \cite{kj} 
as shown in Fig. 3(d). It was shown \cite{n1,n2}  
that there may be increase in distance ($\approx 18\%$) between 
consecutive phosphates which drive backbone to a new torsional 
state \cite{n3}. However $1-2$ kcal/mol energy difference between
two states could not reveal which conformation (c3'-endo or c2'-endo)
poly(dA) will take under equilibrium condition.
In order to see whether the second plateau is a result of 
structural transition, we calculate the number of 
unstacked bases exactly in our model which may participate 
in this transition. 
We increase the reaction coordinate
by a factor proportional to the number of unstacked bases.
The proportionality constant 
for this case has been obtained from Ref. \cite{n3}.
The modified F-x curve depicted in Fig. 4(b)  
shows the multi step plateau which is qualitatively similar to the 
experimental one \cite{ke}.  This implies that further increase in 
the extension (second plateau) is associated with structural transition. 
Moreover, curves for different values 
of $\sigma > 0.51$ almost overlaps with the curve for $\sigma = 1$ 
and hence the transition appears to be weakly cooperative. 

To get the enhanced understanding of the structural changes, we study 
the probability distribution [$P(N_1)$] of poly(dA) from the following 
expression (for $\sigma = 1$): 
$P(N_1)  =  \frac{1}{Z_N} \sum_{N_1,x} C_N(N_1,x)  
e^{\beta N_1 \epsilon} e^{\beta F x}$.
In Fig. 4c, we have shown the probability distribution for stacked 
bases $P(N_1)$ for different values of force at a fixed temperature 
$T=0.3$ for the poly(dA). 
The maxima of P($N_1$) for $F= 0$ occurs at $N_1 \approx 15$. This 
represents the situation where most of the bases are stacked 
(helix) and domains are randomly oriented. 
Slight increase in the force aligns the helix along the force 
direction and thus the number of stacked bases increases slightly. 
This can be seen from Fig. 4c where maxima of $P(N_1)$ shifts 
toward the right. For low forces, thermal fluctuations are too 
weak to unstack the bases in the strand. We observe that there is 
emergence of a new peak around the value $N_1=8$ and decrease 
in the peak heights around $N_1=15$ at the intermediate forces 
($F = 0.2$ to $0.4$). The first plateau took place at $F=0.365$ 
as shown in Fig. 4(b). At this force the height of the both peaks 
are found to be equal. Two peaks of equal height in the probability 
distribution curve shows the signature of co-existence of two 
phases (helix and coil). 
This is analogous to liquid-gas transition. 
Here force represents the pressure and extension is analogous to 
volume \cite{buhot2}. 
In this region the number of stacked bases 
decreases which reflects the unwinding of helix. This gives relative 
extension ($\sim 70\%$) in the backbone.
Further rise in force (above $F=0.5$) shifts the maxima of 
distribution curve toward left. This reflects that the number of 
stacked bases in the stretched state decreases gradually and 
gives additional increase in extension which is about $16\%$ as 
shown in Fig. 4(b). 


In order to resolve the issue associated with structural transition
in poly(dA), it is essential to calculate the change in energy associated 
with this transition. Ke {\it et al} obtained the total energy 
($3.6 \pm 0.2$ kcal/mol per base) from the area under F-x curve 
for adenine (Fig. 2(b) of Ref. \cite{ke}) and interpreted it as base 
stacking energy. In fact total area under the curve [poly(dA)] should be 
attributed to entropic, enthalpic, structural and elastic contribution.
However, in case of F-x curve of thymine [poly(dT)], the contributions come 
from entropic and elastic part only.
Therefore, at low force (neglecting the elastic contribution) 
one can get the rough estimate of enthalpic contribution 
(0.6 kcal/mol per base) associated with coil-helix transition 
corresponding to first plateau by subtracting area under II 
from I below extension 0.58. This value is in agreement with the 
known values available in the literature \cite{goddard}. 
However, at high force, elastic contribution for
thymine is different than the adenine and hence one cannot get
the energy required for structural transition from the same curves.
This may be visualized that area under thymine is more than the
adenine.  By using the same persistence length for adenine \cite{seol1},
one can get F-x curve from modified freely jointed chain (mFJC) which contains only entropic 
and elastic contributions which is shown in Fig. 4. 
The energy ($1.2 \pm 0.2$ kcal/mol per base) associated with 
structural transition above the extension 0.58 has been 
obtained by subtracting the mFJC curve from the experimental F-x 
curve. This value is in excellent agreement with quantum 
calculation \cite{folo}.

We have studied effect of cooperativity in the frame
work of exact enumeration. Our analysis exactly showed that stacking 
reduces the cooperativity [Fig.2(a)] and F-x  curve for $\sigma$ upto 
$0.26$ can be reproduced by stacking alone. For high cooperativity 
($\sigma<0.26$), stacking is not sufficient enough to compensate 
cooperativity factor. 
However, formation of helix reduced the reaction coordinate
which gives an additional factor ($\alpha N_1$) in the partition 
function. We find that this term has significant impact on cooperativity 
which makes plateau broader in case of adenine. Response of 
cooperativity on the applied force below and above the crossover 
point is also evident from Fig. 2(c).

The exact results on short chain of a new semi-microscopic model 
presented here provides unequivocal support for the structural 
transition. This has been substantiated by the correct analysis 
of experimental curve along with the mFJC curve that gives the 
required energy associated with structural transition which is 
in good agreement with quantum calculation. This also resolves 
that c3'-endo puker conformation is most likely in the 
equilibrium condition and at high force poly(dA) goes from c3'-endo 
to c2'-endo conformation.

We would like to thank D. Dhar, S. M. Bhattacharjee and  R. Everaers 
for many helpful discussion. We also thank P. E. Marszalek for 
providing the experimental data and subsequent discussion. 
Financial supports from DST, New Delhi and
UGC, New Delhi are gratefully acknowledged. We thank MPIPKS, 
Dresden for providing the computer resources.


\begin{thebibliography}{99}
\bibitem{n3} It may be noted that adenine has structurally different 
conformations arising due to the rotation of furanose ring in the 
nucleic acids.  The two basic forms $C3'$-endo ($a=5.9 A^o$) and 
$C2'$-endo ($a=7.0 A^o$) of furanose ring characterizes two 
distinctly different families of nucleic acid structures. 
The reorientation of bases at high force may lead to a conformational 
transition in the deoxyribofuranose rings from $C3'$-endo pucker 
to $C2'$-endo pucker in a step like jump which may give about $18 \%$ 
extension in inter phosphate distnace.
\bibitem{ssdna} Y. Seol {\it et al.} have considered the RNA where the number
of bases per helix is different than the ssDNA. We reduce
the reaction coordinate in ssDNA by  the factor $0.35 N_1$.
\end{thebibliography}
\end{document}